\documentclass{article}
\def\Ref#1{(\ref{#1})}
\usepackage{amsmath}
\usepackage{cite}
\newcommand{\rd}{\mathrm{d}}

\newcommand{\st}{\mathrm{st}}
\newcommand{\ho}{\mathrm{hom}}

\newcommand{\fb}{\mathbf{f}}
\begin{document}
\begin{titlepage}
\begin{center}
{\large \textbf{Nonuniform autonomous one-dimensional exclusion
nearest-neighbor reaction-diffusion models}} \vskip 2\baselineskip
\centerline {\sffamily Amir Aghamohammadi\footnote
{mohamadi@alzahra.ac.ir} \& Mohammad Khorrami
\footnote{mamwad@mailaps.org}}
 \vskip 2\baselineskip
{\it Department of Physics, Alzahra University, Tehran 19384,
IRAN}
\end{center}


\begin{abstract}\noindent
The most general nonuniform reaction-diffusion models on a
one-dimensional lattice with boundaries, for which the time
evolution equations of correlation functions are closed, are
considered. A transfer matrix method is used to find the static
solution. It is seen that this transfer matrix can be obtained in
a closed form, if the reaction rates satisfy certain conditions.
We call such models superautonomous. Possible static phase
transitions of such models are investigated. At the end, as an
example of superautonomous models, a nonuniform voter model is
introduced, and solved explicitly.
\end{abstract}
\end{titlepage}
\newpage
\section{Introduction}
Most of the investigations on reaction-diffusion models are
devoted to uniform models, where interaction rates are
site-independent. Among the simplest generalizations beyond a
completely uniform system is a lattice with alternating rates. 
In \cite{GHMS}, relaxation in the kinetic Ising model on an
alternating isotopic chain has been discussed. In
\cite{SchSch02,SchSch03,MObZ}, the steady state configurational
probabilities of an Ising spin chain driven out of equilibrium by
a coupling to two heat baths has been investigated. An example is
a one-dimensional Ising model on a ring, in which the evolution is
according to a generalization of Glauber rates, such that spins at
even (odd) lattice sites experience a temperature $T_{\mathrm{e}}$
($T_{\mathrm{o}}$). In this model the detailed balance is
violated. The response function to an infinitesimal magnetic field
for the Ising-Glauber model with arbitrary exchange couplings has
been studied in \cite{Chatelain}. Other generalizations of the
Glauber model consist of, for example, alternating-isotopic chains
and alternating-bound chains (\cite{GO} for example). In a recent
article \cite{ihg}, we studied the expectation values of spins in
an Ising model with nonuniform coupling constants. A transfer
matrix method was used to study the steady state behavior of the
system in the thermodynamic limit. Different (static) phases of
this system were studied, and a closed form was obtained for this
transfer matrix.

In \cite{26} a ten-parameter family of one-species
reaction-diffusion processes with nearest-neighbor interaction was
introduced, for which the evolution equation of $n$-point
functions contains only $n$- or less- point functions, the so
called autonomous models. The average particle-number in each site
was obtained exactly for these models. In \cite{27,28}, this was
generalized to multi-species systems and more-than-two-site
interactions. In \cite{29,30,32}, the phase structure of some classes
of single or multiple-species reaction-diffusion systems was
investigated. These investigations were based on the one-point
functions of the systems.

In the present paper the most general nonuniform exclusion
nearest-neighbor reaction-diffusion models on a one-dimensional
lattice with boundaries are studied, for which the evolution
equations of the one-point functions are closed, and the transfer
matrix has a closed form. It is shown that there is a possible
phase transition in such models, which corresponds to a reduction of
the role of boundary conditions on time-independent profile of
the expectation value of the number operators. The scheme of the
paper is as follows. In section 2, the models are introduced, and
the evolution equation for the expectation values of $n_i$ (the
number operators at the site $i$) is obtained. Also conditions are
obtained so that the evolution of the expectation values of $n_i$ is
closed. In section 3, the equation governing the static solution for
the expectation values of $n_i$'s is obtained, and a transfer matrix
method is introduced to obtain the static solution and investigate
different (static) phases of the system. It is also seen that to
write a closed form for the transfer matrix, further conditions on
the reaction rates are to be satisfied. We call models satisfying
these conditions superautonomous models. In section 4, as an
example, a nonuniform voter model is investigated in more detail.
Section 5 is devoted to the concluding remarks.
\section{Exclusion nearest-neighbor reaction-diffusion models with nonuniform reaction rates}
Consider a one-dimensional lattice with $(L+1)$ sites, numbered
from $0$ to $L$. Each site is either empty (denoted by the vector $e_0$) or occupied with one
particle (denoted by the vector $e-1$). The evolution of the system is said to be governed by
nearest-neighbor interactions, if the evolution of each site depends on only that site and its
nearest neighbors (sites directly related to it through a link). The evolution of such a system is
governed by a Hamiltonian
${\mathcal H}$ of the form,
\begin{equation}\label{nu.01}
{\mathcal H} ={\mathcal H}'_0+\left(\sum_\alpha{\mathcal
H}_\alpha\right)+{\mathcal H}'_L,
\end{equation}
where ${\mathcal H}_\alpha$ corresponds to the link $\alpha$:
\begin{equation}\label{nu.02}
{\mathcal H}_\alpha=1^{\otimes(\alpha-\mu)}\otimes
H_\alpha\otimes 1^{\otimes(L-\alpha-\mu)},
\end{equation}
and
\begin{equation}\label{nu.03}
\mu:=\frac{1}{2}.
\end{equation}
The link $\alpha$ links the sites $(\alpha-\mu)$ and
$(\alpha+\mu)$, so that $\alpha\pm\mu$ are integers, and $\alpha$
runs from $\mu$ up to $(L-\mu)$. Throughout this paper, sites are
denoted by Latin letters which represent integers, while links are
denoted by Greek letters which represent integers plus one half
($\mu$), so that the link $\alpha$ joins the sites $(\alpha-\mu)$
and $(\alpha+\mu)$, while the site $i$ joins the links $(i-\mu)$
and $(i+\mu)$. $H_\alpha$ is a linear operator acting on a four
dimensional space (the configuration space corresponding to the
sites $(\alpha-\mu)$ and $(\alpha+\mu)$) with a basis
$\{e_{0\,0},e_{0\,1},e_{1\,0},e_{1\,1}\}$. Also,
\begin{align}\label{nu.04}
{\mathcal H}'_0&=
H'_0\otimes 1^{\otimes L}, \nonumber\\
{\mathcal H}'_L&=1^{\otimes L}\otimes H'_L,
\end{align}
where $H'0$ and $H'_L$ are linear operators acting on two
dimensional spaces (the configuration spaces corresponding to the
sites $0$ and $L$, respectively) with bases $\{e_0,e_1\}$.
The nondiagonal components of $H_\alpha$, $H'_0$, and $H'_L$ are reaction rates. Denoting a full
site by $\bullet$ and an empty site by $\circ$, the possible reactions for the boundary sites $0$ or $L$
are
\begin{align}\label{nu.05}
\circ\to\bullet,\quad\hbox{with the rate }&(H'_{0,L})^1{}_0,\nonumber\\
\bullet\to\circ,\quad\hbox{with the rate }&(H'_{0,L})^0{}_1,
\end{align}
while those for the link $\alpha$ are
\begin{align}\label{nu.06}
\circ\circ\to\circ\bullet,\quad\hbox{with the rate }&(H_\alpha)^{01}{}_{00},\nonumber\\
\circ\circ\to\bullet\circ,\quad\hbox{with the rate }&(H_\alpha)^{10}{}_{00},\nonumber\\
\circ\circ\to\bullet\bullet,\quad\hbox{with the rate }&(H_\alpha)^{11}{}_{00},\nonumber\\
\circ\bullet\to\circ\circ,\quad\hbox{with the rate }&(H_\alpha)^{00}{}_{01},\nonumber\\
\circ\bullet\to\bullet\circ,\quad\hbox{with the rate }&(H_\alpha)^{10}{}_{01},\nonumber\\
\circ\bullet\to\bullet\bullet,\quad\hbox{with the rate }&(H_\alpha)^{11}{}_{01},\nonumber\\
\bullet\circ\to\circ\circ,\quad\hbox{with the rate }&(H_\alpha)^{00}{}_{10},\nonumber\\
\bullet\circ\to\circ\bullet,\quad\hbox{with the rate }&(H_\alpha)^{01}{}_{10},\nonumber\\
\bullet\circ\to\bullet\bullet,\quad\hbox{with the rate }&(H_\alpha)^{11}{}_{10},\nonumber\\
\bullet\bullet\to\circ\circ,\quad\hbox{with the rate }&(H_\alpha)^{00}{}_{11},\nonumber\\
\bullet\bullet\to\circ\bullet,\quad\hbox{with the rate }&(H_\alpha)^{01}{}_{11},\nonumber\\
\bullet\bullet\to\bullet\circ,\quad\hbox{with the rate }&(H_\alpha)^{10}{}_{11}.
\end{align}
It is seen that these rates are in general different for different links (also the rates on the boundary sites are in general different). The system is called uniform if the rates are the same for all links,
and nonuniform if it is not the case.

The number operator in the site $i$ is denoted by $n_i$:
\begin{equation}\label{nu.07}
n_i=1^{\otimes i}\otimes n
\otimes 1^{\otimes(L-i)},
\end{equation}
where $n$ is an operator acting on a two dimensional space with the
basis $\{e_0,e_1\}$. The matrix form of $n$ in this basis is
\begin{equation}\label{nu.08}
n^a{}_b=\delta^a_1\,\delta^1_b.
\end{equation}

The evolution equation for the expectation value of an observable
$Q$ is
\begin{equation}\label{nu.09}
\frac{\rd}{\rd t}\langle Q\rangle=\langle Q\,{\mathcal H}\rangle,
\end{equation}
where
\begin{equation}\label{nu.10}
\langle Q\rangle=S\,Q\,\Psi,
\end{equation}
the vector $\Psi$ is the ($2^{L+1}$ dimensional) probability vector
describing the system and $S$ is the covector
\begin{equation}\label{nu.11}
S:=s^{\otimes(L+1)},
\end{equation}
and
\begin{equation}\label{nu.12}
s_a=1.
\end{equation}

The system is called autonomous, if the Hamiltonian is so that
the evolution of the expectation values of $n_i$ is closed in
terms of the expectation values of $n_j$'s. In the evolution
equation for the expectation value of $n_i$, the expectation
values of $n_{i-1}$, $n_i$, $n_{i+1}$, $(n_{i-1}\,n_i)$, and
$(n_i\,n_{i+1})$ occur. It is seen that the criterion that the
coefficients of the last two vanish, is
\begin{align}\label{nu.13}
s_a\,[H_{i-\mu}\,r\otimes\,r]^{a\,1}=&
0,\nonumber\\
s_a\,[H_{i+\mu}\,r\otimes\,r]^{1\,a}=& 0,
\end{align}
respectively, where
\begin{equation}\label{nu.14}
r^a=-\delta^a_0+\delta^a_1.
\end{equation}
Equation \Ref{nu.13} should hold for all $i$'s, in order that the system be
autonomous. So one can rewrite it like
\begin{align}\label{nu.15}
s_a\,[H_{\alpha}\,r\otimes\,r]^{a\,1}&=0,\nonumber\\
s_a\,[H_{\alpha}\,r\otimes\,r]^{1\,a}&= 0.
\end{align}
It is seen that this condition is the same as the corresponding
condition for uniform lattices \cite{26,30}, written for each link separately.
Provided that this condition holds, one arrives at
\begin{equation}\label{nu.16}
\!\frac{\rd}{\rd t}\langle n_i\rangle=\eta_{i-\mu}\,\langle
n_{i-1}\rangle+\theta_{i+\mu}\,\langle n_{i+1}\rangle
+(\kappa_{i-\mu}+\nu_{i+\mu})\,\langle n_i\rangle
+(\xi_{i-\mu}+\sigma_{i+\mu}), \!\!\!\quad 0<i<L,
\end{equation}
where
\begin{align}\label{nu.17}
\eta_\alpha&:=s_a\,(H_\alpha)^{a\,1}{}_{b\,0}\,r^b,\nonumber\\
\theta_\alpha&:=s_a\,(H_\alpha)^{1\,a}{}_{0\,b}\,r^b,\nonumber\\
\kappa_\alpha&:=s_a\,(H_\alpha)^{a\,1}{}_{0\,b}\,r^b,\nonumber\\
\nu_\alpha&:=s_a\,(H_\alpha)^{1\,a}{}_{b\,0}\,r^b,\nonumber\\
\xi_\alpha&:=s_a\,(H_\alpha)^{a\,1}{}_{0\,0},\nonumber\\
\sigma_\alpha&:=s_a\,(H_\alpha)^{1\,a}{}_{0\,0}.
\end{align}

For the boundary sites (the sites $0$ and $L$), one has
\begin{align}\label{nu.18}
\frac{\rd}{\rd t}\langle n_0\rangle&=\theta_\mu\,\langle
n_1\rangle +[(H'_0)^1{}_1-(H'_0)^0{}_1+\nu_\mu]\,\langle
n_0\rangle +[(H'_0)^0{}_1+\sigma_\mu],\\ \label{nu.19}
\frac{\rd}{\rd t}\langle n_L\rangle&=\eta_{L-\mu}\,\langle
n_{L-1}\rangle
+[\kappa_{L-\mu}+(H'_L)^1{}_1-(H'_L)^0{}_1]\,\langle n_L\rangle
+[\xi_{L-\mu}+(H'_L)^0{}_1].
\end{align}

\section{The static solution}
For the static solution ($\langle n\rangle_\st$), the left hand
side of \Ref{nu.16} vanishes and one obtains
\begin{equation}\label{nu.20}
\langle n_{i+1}\rangle_\st=
-\frac{\eta_{i-\mu}}{\theta_{i+\mu}}\,\langle n_{i-1}\rangle_\st
-\frac{\kappa_{i-\mu}+\nu_{i+\mu}}{\theta_{i+\mu}}\,\langle
n_i\rangle_\st -\frac{\xi_{i-\mu}+\sigma_{i+\mu}}{\theta_{i+\mu}}.
\end{equation}
Denoting that part of this solution which satisfies the
homogeneous equation by $\langle n\rangle_\st^\ho$, it is seen
that
\begin{equation}\label{nu.21}
\langle n_{i+1}\rangle_\st^\ho=
-\frac{\eta_{i-\mu}}{\theta_{i+\mu}}\,\langle
n_{i-1}\rangle_\st^\ho
-\frac{\kappa_{i-\mu}+\nu_{i+\mu}}{\theta_{i+\mu}}\,\langle
n_i\rangle_\st^\ho,
\end{equation}
which  can be written as the following matrix form
\begin{equation}\label{nu.22}
X_{i+\mu}^\ho=D_i\,X_{i-\mu}^\ho,
\end{equation}
where
\begin{equation}\label{nu.23}
X_\alpha:=\begin{bmatrix}\langle n_{\alpha-\mu}\rangle_\st\\
\langle n_{\alpha+\mu}\rangle_\st\end{bmatrix},
\end{equation}
and
\begin{equation}\label{nu.24}
D_i:=\begin{bmatrix}0&1 \\ & \\
\displaystyle{-\frac{\eta_{i-\mu}}{\theta_{i+\mu}}}&
\displaystyle{-\frac{\kappa_{i-\mu}+\nu_{i+\mu}}{\theta_{i+\mu}}}\end{bmatrix}.
\end{equation}
Using these, one arrives at
\begin{equation}\label{nu.25}
X_\alpha^\ho=D_{\alpha\,\beta}\,X_\beta^\ho,
\end{equation}
where
\begin{equation}\label{nu.26}
D_{\alpha\,\beta}:=D_{\alpha-\mu}\,D_{\alpha-\mu-1}\cdots
D_{\beta+\mu}.
\end{equation}
To solve \Ref{nu.20}, one can use a tranfer matrix (Green's
function) method. Consider the equation
\begin{equation}\label{nu.27}
G_{i+1\,j}= -\frac{\eta_{i-\mu}}{\theta_{i+\mu}}\,G_{i-1\, j}
-\frac{\kappa_{i-\mu}+\nu_{i+\mu}}{\theta_{i+\mu}}\,G_{i\,j}
-\delta_{i\,j}.
\end{equation}
Defining
\begin{equation}\label{nu.28}
Y_{\alpha\,j}:=\begin{bmatrix}G_{\alpha-\mu\,j}\\
G_{\alpha+\mu\,j}\end{bmatrix},
\end{equation}
it is seen that the solution for $Y$ is
\begin{equation}\label{nu.29}
Y_{\alpha\,j}=\begin{cases} D_{\alpha\,\beta}\,\tilde
Y_{\beta\,j},& \alpha<j\\
D_{\alpha\,\beta}\,Y_{\beta\,j},& \alpha>j
\end{cases},
\end{equation}
with the condition that \Ref{nu.27} holds for $i=j$. This
condition is
\begin{equation}\label{nu.30}
Y_{j+\mu\,j}=D_j\,Y_{j-\mu\,j}-\begin{bmatrix}0\\
1\end{bmatrix},
\end{equation}
which reads
\begin{equation}\label{nu.31}
D_{j+\mu\,\beta}\,(Y_{\beta\,j}-\tilde Y_{\beta\,j})=-\begin{bmatrix}0\\
1\end{bmatrix}.
\end{equation}
A particular solution for $Y$ is obtained if one sets $\tilde Y$
equal to zero. In this case,
\begin{equation}\label{nu.32}
Y_{\alpha\,j}=-\Theta_{\alpha\,j}\,D_{\alpha\,j+\mu}\,\begin{bmatrix}0\\
1\end{bmatrix},
\end{equation}
where $\Theta$ is the step function:
\begin{equation}\label{nu.33}
\Theta_{\alpha\,j}:=\begin{cases} 0,& \alpha<j\\
1,& \alpha>j
\end{cases}.
\end{equation}
Using this, the general solution to \Ref{nu.20} can be written as
\begin{equation}\label{nu.34}
X_\alpha=D_{\alpha\,\beta}\,X_\beta^\ho-
\sum_{j<\alpha}D_{\alpha\,j+\mu}
\frac{\xi_{j-\mu}+\sigma_{j+\mu}}{\theta_{j+\mu}}
\begin{bmatrix}0\\1\end{bmatrix}.
\end{equation}

As it was the case in \cite{ihg}, the steady state profile near
the end-site $0$ is determined by the eigenvalues of the matrix
$D_{\alpha\,\mu}$, where $\alpha$ is some site far from the ends.
One has
\begin{align}\label{nu.35}
X_\alpha&=X_\alpha^a\,\fb_a,\nonumber\\
X_{\mu}&=X_{\mu}^a\,\fb_a,
\end{align}
where $\fb_a$ is the eigenvector of $D_{\alpha\,\mu}$
corresponding to the eigenvalue $\lambda^a$, and $X_\alpha^a$'s
and $X_{\mu}^a$'s are the coefficients of expansions of $X_\alpha$
and $X_{\mu}$ in terms of the eigenvectors. If there was no
nonhomogeneous part in the equation \Ref{nu.16}, then the
discussion would be exactly similar to that of \cite{ihg}:
$X_{\mu}^a$ vanishes if $\lambda^a$ tends to infinity (in the
thermodynamic limit). Otherwise, $X_{\mu}^a$ is generally nonzero
and determined by the boundary conditions. If the nonhomogeneous
part does not vanish, then the second term in \Ref{nu.34}
generally contains a large multiple of $\fb_a$ if $\lambda^a$
tends to infinity. This large part is to be cancelled by a large
multiple of $\fb_a$ coming from the first term in \Ref{nu.34}. So
in this case $X_{\mu}^a$ does not vanish but tends to a fixed
value independent of boundary conditions. It is seen that although
equations \Ref{nu.18} and \Ref{nu.19} serve as boundary conditions
to obtain say $X_\mu$, the above general argument is independent
of these conditions. The essence of the above argument is the following.
In general, $X_\alpha$ (the components of which are noting but
expectations of number operators) is a linear combination of two
vectors ($\fb_a$'s) plus a nonhomogeneous part. The nonhomogeneous
part is determined from the bulk reactions, and the two unknown
coefficients $\fb_a$'s are to be determined from the the boundary
conditions resulted from \Ref{nu.18} and \Ref{nu.19} (in the static
case that the left hand sides vanish). There are, however, regions
of the parameter space where in the thermodynamic limit one of the
coefficients of $\fb_a$'s (or possibly both) are determined from
the bulk reactions only. This essentially means that the effect of
boundaries on the behavior of the system is reduced. In a transport
system, for example, it is expected that the time-independent profile
of the moving bodies' density depend on both the bulk reactions
(speed, overtaking, etc) and the boundary reactions (injection and
extraction rates). But there could be cases where these boundary
terms are unimportant, or less important.

The situation (in the thermodynamic limit) can be summarized as
follows.
\begin{itemize}
\item[\textbf{i}] The eigenvalue $\lambda^a$ tends to infinity.
In this case $X_{\mu}^a$ tends to a fixed value, independent of
the boundary conditions.
\item[\textbf{ii}] The eigenvalue $\lambda^a$ tends to zero or a
finite number. In this case $X_{\mu}^a$ is determined by the
boundary condiotions.
\end{itemize}
Obviously, similar cases occur at the other boundary site. It is seen
that this behavior at one of the boundaries is independent of the
analog behavior at the other boundary.

This is the static phase transition of the system, some discontinuous behavior of
the expectation value of the number operator near the boundaries (note that the
components of $X_\alpha$ are nothing but the expectations of number operators).

If one can write $D_i$ as
\begin{equation}\label{nu.36}
D_i:=\Sigma_{i+\mu}\,\Delta_i\,\Sigma^{-1}_{i-\mu},
\end{equation}
where $\Delta_i$ is diagonal, and $\Sigma_\alpha$ depends on only
the parameters corresponding to the link $\alpha$, then it is easy
to find the solution to \Ref{nu.22}. The system is called
superautonomous, if this is the case. Putting
\begin{equation}\label{nu.37}
\Sigma_\alpha=\begin{bmatrix}a_\alpha &b_\alpha \\
c_\alpha &d_\alpha\end{bmatrix},
\end{equation}
and
\begin{equation}\label{nu.38}
\Delta_i=\begin{bmatrix}
A_i& 0 \\ & \\
0 & B_i\end{bmatrix},
\end{equation}
in \Ref{nu.36}, one arrives at
\begin{align}\label{nu.39}
\frac{A_i}{\varsigma_{i-\mu}}\,a_{i+\mu}\,d_{i-\mu}
-\frac{B_i}{\varsigma_{i-\mu}}\,b_{i+\mu}\,c_{i-\mu}&=0,\\
\label{nu.40} -\frac{A_i}{\varsigma_{i-\mu}}\,a_{i+\mu}\,b_{i-\mu}
+\frac{B_i}{\varsigma_{i-\mu}}\,b_{i+\mu}\,a_{i-\mu}&=1,\\
\label{nu.41} \frac{A_i}{\varsigma_{i-\mu}}\,c_{i+\mu}\,d_{i-\mu}
-\frac{B_i}{\varsigma_{i-\mu}}\,d_{i+\mu}\,c_{i-\mu}&=
-\frac{\eta_{i-\mu}}{\theta_{i+\mu}},\\
\label{nu.42} -\frac{A_i}{\varsigma_{i-\mu}}\,c_{i+\mu}\,b_{i-\mu}
+\frac{B_i}{\varsigma_{i-\mu}}\,d_{i+\mu}\,a_{i-\mu}&=
-\frac{\kappa_{i-\mu}+\nu_{i+\mu}}{\theta_{i+\mu}},
\end{align}
where
\begin{equation}\label{nu.43}
\varsigma_\alpha:=a_\alpha\,d_\alpha-b_\alpha\,c_\alpha.
\end{equation}
Using \Ref{nu.39} and \Ref{nu.40}, one obtains $A_i$ and $B_i$:
\begin{align}\label{nu.44}
A_i&=\frac{c_{i-\mu}}{a_{i+\mu}},\\
\label{nu.45} B_i&=\frac{d_{i-\mu}}{b_{i+\mu}}.
\end{align}
Using these in \Ref{nu.41}, it is seen that
\begin{equation}\label{nu.46}
\frac{c_{i-\mu}\,d_{i-\mu}}{\varsigma_{i-\mu}}\,
\frac{\varsigma_{i+\mu}}{a_{i+\mu}\,d_{i+\mu}}=
\frac{\eta_{i-\mu}}{\theta_{i+\mu}},
\end{equation}
which can be solved as
\begin{align}\label{nu.47}
a_\alpha\,b_\alpha&=\phi\,\theta_\alpha\,\varsigma_\alpha,\\
\label{nu.48}
c_\alpha\,d_\alpha&=\phi\,\eta_\alpha\,\varsigma_\alpha,
\end{align}
where $\phi$ is a constant. Finally, using \Ref{nu.44},
\Ref{nu.45}, \Ref{nu.47}, and \Ref{nu.48} in \Ref{nu.42}, on obtains
\begin{equation}\label{nu.49}
\frac{a_{i-\mu}\,d_{i-\mu}}{\varsigma_{i-\mu}}
+\frac{b_{i+\mu}\,c_{i+\mu}}{\varsigma_{i+\mu}}=
-\phi\,\kappa_{i-\mu}-\phi\,\nu_{i+\mu},
\end{equation}
the solution to which is
\begin{align}\label{nu.50}
a_\alpha\,d_\alpha&=(-\phi\,\kappa_\alpha+\psi)\,\varsigma_\alpha,\\
\label{nu.51}
b_\alpha\,c_\alpha&=(-\phi\,\nu_\alpha-\psi)\,\varsigma_\alpha,
\end{align}
where $\psi$ is another constant. $\varsigma$ is not an
independent variable in $\Sigma$. Putting \Ref{nu.50} and
\Ref{nu.51} in \Ref{nu.43}, one arrives at
\begin{equation}\label{nu.52}
2\,\psi=\phi\,(\kappa_\alpha-\nu_\alpha)+1,
\end{equation}
which shows that $(\kappa-\nu)$ should be constant. One can use
\Ref{nu.52} in \Ref{nu.50} and \Ref{nu.51}, to obtain
\begin{align}\label{nu.53}
a_\alpha\,d_\alpha&=\frac{\varsigma_\alpha}{2}\,
[-\phi\,(\kappa_\alpha+\nu_\alpha)+1],\\
\label{nu.54} b_\alpha\,c_\alpha&=\frac{\varsigma_\alpha}{2}\,
[-\phi\,(\kappa_\alpha+\nu_\alpha)-1].
\end{align}
These two equations are not independent of \Ref{nu.47} and
\Ref{nu.48}. The consistency condition is
\begin{equation}\label{nu.55}
\phi^2(\kappa_\alpha+\nu_\alpha)^2-1=4\,\phi^2\,\eta_\alpha\,\theta_\alpha,
\end{equation}
showing that $[(\kappa+\nu)^2-4\,\eta\,\theta]$ should be
constant. Noting that $(\kappa-\nu)$ has to be a constant as well,
the second condition can be stated as $(\kappa\,\nu-\eta\,\theta)$
should be constant. So the conditions that the system be
superautonomous are
\begin{align}\label{nu.56}
\kappa_\alpha-\nu_\alpha&=\mbox{constant},\\ \label{nu.57}
\kappa_\alpha\,\nu_\alpha-\eta_\alpha\,\theta_\alpha&=\mbox{constant}.
\end{align}

There are some special cases resembling those encountered in
\cite{ihg}.
\begin{itemize}
\item[\textbf{1}] Constant coupling:
Here $H_\alpha$ does not depend on $\alpha$, and $\lambda^a$ tends
to infinity (zero) if and only if the corresponding eigenvalue of $D_i$ is
greater (smaller) than one.
\item[\textbf{2}] Periodic coupling:
\begin{equation}\label{nu.58}
H_{\alpha+m}=H_\alpha.
\end{equation}
In this case the behavior of the eigenvalues of $D_{\alpha,\mu}$
is determined by the eigenvalues of $D_{\alpha+m,\alpha}$: An
eigenvalue of $D_{\alpha,\mu}$ tends to infinity (zero) if and only if the
corresponding eigenvalue of $D_{\alpha+m,\alpha}$ is greater
(smaller) than one.
\item[\textbf{3}] Defects in the lattice:
No new phenomena is seen, as long as the defects are localized,
i.e. they are far from the boundaries. So if there is a lattice
that has some defects but otherwise is uniform, the static
behavior near the boundaries is similar to that of a uniform
lattice \cite{30}.
\item[\textbf{4}] A lattice with different behaviors at different end points:
The behaviors of the static solution near the two ends are
independent of each other, provided the behavior change occurs far
from the boundaries. So all the phenomena seen in previous special
cases can be seen at each boundary, independent of the other boundary.
\end{itemize}
\section{An example, the voter model}
The voter model is a lattice each site of which is either full
($\bullet$) or empty ($\circ$). The reactions on a link are
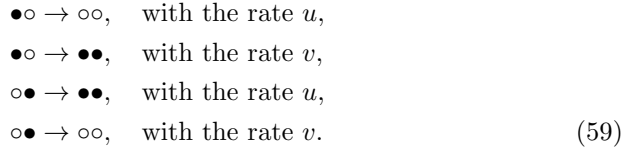
\begin{align}\label{nu.59}
\bullet\circ\to\circ\circ,\quad\hbox{with the rate }&u,\nonumber\\
\bullet\circ\to\bullet\bullet,\quad\hbox{with the rate }&v,\nonumber\\
\circ\bullet\to\bullet\bullet,\quad\hbox{with the rate }&u,\nonumber\\
\circ\bullet\to\circ\circ,\quad\hbox{with the rate }&v.
\end{align}
Of course, the rates $u$ and $v$ may be link-dependent. using
\Ref{nu.17},
\begin{align}\label{nu.60}
\eta_\alpha&=v_\alpha,\nonumber\\
\theta_\alpha&=u_\alpha,\nonumber\\
\kappa_\alpha&=-v_\alpha,\nonumber\\
\nu_\alpha&=-u_\alpha,\nonumber\\
\xi_\alpha&=0,\nonumber\\
\sigma_\alpha&=0.
\end{align}
In order that the system be superautonomous, \Ref{nu.56} and
\Ref{nu.57} must hold. \Ref{nu.56} reads
\begin{equation}\label{nu.61}
u_\alpha-v_\alpha=\mbox{constant},
\end{equation}
and \Ref{nu.57} is an identity. Assume that \Ref{nu.61} holds.
Using \Ref{nu.24}, one has
\begin{equation}\label{nu.62}
D_i:=\begin{bmatrix}0&1 \\ & \\
\displaystyle{-\frac{v_{i-\mu}}{u_{i+\mu}}}&
\displaystyle{1+\frac{v_{i-\mu}}{u_{i+\mu}}}\end{bmatrix}.
\end{equation}
From \Ref{nu.55}, one has
\begin{equation}\label{nu.63}
\phi^2\,(u_\alpha-v_\alpha)^2=1,
\end{equation}
one of the solutions to which is
\begin{equation}\label{nu.64}
\phi=\frac{1}{u_\alpha-v_\alpha}.
\end{equation}
Putting this in \Ref{nu.47}, \Ref{nu.48}, \Ref{nu.53}, and
\Ref{nu.54}, one arrives at
\begin{align}\label{nu.65}
a_\alpha\,b_\alpha&=\phi\,u_\alpha\,\varsigma_\alpha,\\
\label{nu.66} c_\alpha\,d_\alpha&=\phi\,v_\alpha\,\varsigma_\alpha,\\
\label{nu.67} a_\alpha\,d_\alpha&=\phi\,u_\alpha\,\varsigma_\alpha,\\
\label{nu.68}
b_\alpha\,c_\alpha&=\phi\,v_\alpha\,\varsigma_\alpha.
\end{align}
One set of solutions to these equations is
\begin{equation}\label{nu.69}
\Sigma_\alpha=\begin{bmatrix}u_\alpha &1 \\
v_\alpha &1\end{bmatrix}.
\end{equation}
Putting this in \Ref{nu.44} and \Ref{nu.45} results in
\begin{equation}\label{nu.70}
\Delta_i=\begin{bmatrix}
\displaystyle{\frac{v_{i-\mu}}{u_{i+\mu}}}& 0 \\ & \\
0 & 1\end{bmatrix}.
\end{equation}
One also has
\begin{align}\label{nu.71}
\varsigma_\alpha&=u_\alpha-v_\alpha,\nonumber\\
                &=\phi^{-1}.
\end{align}
So
\begin{equation}\label{nu.72}
D_{\alpha\,\mu}=\begin{bmatrix}\displaystyle{\phi\,v_\mu\,
\left(\frac{v_{\alpha-1}\cdots v_{\mu+1}} {u_{\alpha-1}\cdots
u_{\mu+1}}-1\right)}& \displaystyle{1-\phi\,v_\mu\,
\left(\frac{v_{\alpha-1}\cdots v_{\mu+1}} {u_{\alpha-1}\cdots
u_{\mu+1}}-1\right)}\\
\displaystyle{\phi\,v_\mu\, \left(\frac{v_\alpha\cdots v_{\mu+1}}
{u_\alpha\cdots u_{\mu+1}}-1\right)}&
\displaystyle{1-\phi\,v_\mu\, \left(\frac{v_\alpha\cdots
v_{\mu+1}} {u_\alpha\cdots u_{\mu+1}}-1\right)}
\end{bmatrix}.
\end{equation}
One has
\begin{align}\label{nu.73}
\det(D_{\alpha\,\mu})&=\frac{v_\mu}{u_\alpha}\,
\left(\frac{v_{\alpha-1}\cdots v_{\mu+1}}{u_{\alpha-1}\cdots
u_{\mu+1}}\right),\\ \label{nu.74}
\mathrm{tr}(D_{\alpha\,\mu})&=1+\frac{v_\mu}{u_\alpha}\,
\left(\frac{v_{\alpha-1}\cdots v_{\mu+1}}{u_{\alpha-1}\cdots
u_{\mu+1}}\right),
\end{align}
showing that the eigenvalues of $D_{\alpha\,\mu}$ are
\begin{align}\label{nu.75}
\lambda^1&=1,\\ \label{nu.76} \lambda^2&=\frac{v_\mu}{u_\alpha}\,
\left(\frac{v_{\alpha-1}\cdots v_{\mu+1}}{u_{\alpha-1}\cdots
u_{\mu+1}}\right).
\end{align}

A special case is when
\begin{equation}\label{nu.77}
v_\alpha=u_\alpha.
\end{equation}
In this case the matrix $\Sigma_\alpha$ is singular. Yet one can
obtain the matrix $D_{\alpha\,\mu}$ as a limit $\phi\to\infty$ of
\Ref{nu.72}. The result is
\begin{equation}\label{nu.78}
D_{\alpha\,\mu}=\begin{bmatrix}\displaystyle{-v_\mu\,
\left(\frac{1}{v_{\alpha-1}}+\cdots+\frac{1}{v_{\mu+1}}\right)}&
\displaystyle{1+v_\mu\,
\left(\frac{1}{v_{\alpha-1}}+\cdots+\frac{1}{v_{\mu+1}}\right)}
\\
\displaystyle{-v_\mu\,
\left(\frac{1}{v_\alpha}+\cdots+\frac{1}{v_{\mu+1}}\right)}&
\displaystyle{1+v_\mu\,
\left(\frac{1}{v_\alpha}+\cdots+\frac{1}{v_{\mu+1}}\right)}
\end{bmatrix},
\end{equation}
and both eigenvalues of $D_{\alpha\,\mu}$ become one.

Let's study some special cases.
\begin{itemize}
\item[\textbf{1}] Constant coupling:
\begin{equation}\label{nu.79}
u_\alpha=\mathrm{constant}.
\end{equation}
In this case $\lambda^2$ tends to infinity if and only if $v_\alpha$ is
greater than $u_\alpha$.
\item[\textbf{2}] Periodic coupling:
\begin{equation}\label{nu.80}
u_{\alpha+m}=u_\alpha.
\end{equation}
In this case the behavior of the eigenvalues of $D_{\alpha,\mu}$
is determined by the eigenvalues of $D_{\alpha+m,\alpha}$, which
are one and
\begin{equation}\label{nu.81}
\Lambda^2=\frac{v_{\mu+m-1}\cdots v_{\mu}}{u_{\mu+m-1}\cdots
u_{\mu}}.
\end{equation}
If $u_\alpha$ is greater than $v_\alpha$, then $\lambda^2$ tends
to zero. If $u_\alpha$ is smaller than $v_\alpha$, then
$\lambda^2$ tends to infinity.
\item[\textbf{3}] closed lattice:
Let the $L$'th site be the same as the $0$'th site. One has
\begin{equation}\label{nu.82}
x_{L+\mu}=x_\mu.
\end{equation}
Combining this with \Ref{nu.34}, it is seen that $x_\mu$ should be
the eigenvector of $D_{L+\mu,\mu}$ corresponding to the eigenvalue
one, which shows that
\begin{equation}\label{nu.83}
x_\mu=\begin{bmatrix} 1\\1
\end{bmatrix},
\end{equation}
showing that the stationary profile of the density is uniform.
This is despite the fact that the reaction rates are not
necessarily uniform.
\end{itemize}

\section{Concluding remarks}
General autonomous exclusion models with nearest-neighbor
interactions on a one-dimensional lattice were studied, for them
the reaction rates were nonuniform. By autonomous is is meant that
the evolution equation for the expectation values of the number
operators are closed. It was seen that the condition that the
system be autonomous is the same as the analogous condition for
uniform lattices. A transfer matrix method was introduced to solve
the equation for the static configuration of the expectation
values of the number operators. The static phase picture of these
systems, including possible phase transitions, was investigated.
These phase transitions correspond to a reduction of the role of
boundary reactions on the profile of the expectations of the
number operators, similar to the case of uniform lattices.
Also similar to the case of uniform lattices, these possible
phase transitions are not affected by the boundary conditions.
Moreover, all of the above mentioned possible static phase transitions
near one boundary, are controlled by only the bulk reaction rates
in a large part of the lattice one boundary of which is the same
boundary. So phase transitions at different boundaries are
independent of each other, and finite defects far from boundaries
have no effect on the phase transitions.
It was seen that if the reaction rates satisfy ceratin
additional conditions (which are essentially the constancy of
ceratin combinations of reaction rates) a closed form can be
obtained for the transfer matrix. Systems satisfying those
conditions were called superautonomous. The example of the voter
model was studied in more detail.\\
\\
\textbf{Acknowledgement}:  This work was partially supported by
the research council of the Alzahra University.


\newpage
\begin{thebibliography}{99}
\bibitem{GHMS}      L.~L.~Gon\c{c}alves, M.~L\'{o}pez~de~Haro, J.~Tag\"{u}e\~{n}a-Mart\'{i}nez, \&
                    R.~B.~Stinchcombe; arXiv:cond-mat/9911225
\bibitem{SchSch02}  B.~Schmittmann \&  F.~Schm\"{u}ser;
                    Phys. Rev. \textbf{E66} (2002) 046130.
\bibitem{SchSch03}  B.~Schmittmann \& F.~Schm\"{u}ser;
                    J. Phys. \textbf{A35} (2002) 2569.
\bibitem{MObZ}      M.~Mobilia, R.~K.~P.~Zia, \& B.~Schmittmann;
                    J. Phys. \textbf{A37} (2004) L407.
\bibitem{Chatelain} C.~Chatelain; J. Phys. \textbf{A36} (2003) 10739.
\bibitem{GO}        L.~L.~Gon\c{c}alves \& A.~L.~Stella;
                    J. Phys. \textbf{A20} (1987) L387.
\bibitem{ihg}       M.~Khorrami \& A.~Aghamohammadi; arxiv:0811.2283.
\bibitem{26}        G.~M.~Sch\"{u}tz;  J. Stat. Phys. \textbf{79} (1995) 243.
\bibitem{27}        A.~Aghamohammadi, A.~H.~Fatollahi, M.~Khorrami, \&
                    A.~Shariati; Phys. Rev. \textbf{E62} (2000) 4642.
\bibitem{28}        A.~Shariati, A.~Aghamohammadi, \& M.~Khorrami;
                    Phys. Rev. \textbf{E64} (2001) 066102.
\bibitem{29}        M.~Khorrami \& A.~Aghamohammadi; Phys. Rev. \textbf{E63}
                    (2001) 042102.
\bibitem{30}        A.~Aghamohammadi \& M.~Khorrami; J. Phys. \textbf{A34}
                    (2001) 7431.
\bibitem{32}        M.~Khorrami \& A.~Aghamohammadi; Phys. Rev. \textbf{E65}
                    (2002) 056129.
\end{thebibliography}
\end{document}